\documentclass[11pt]{article}
\usepackage{geometry}
\usepackage{xcolor}
\usepackage{graphicx}
\usepackage{epsf}
\usepackage{amsmath}
\usepackage{amssymb}
\usepackage{cite}
\newcommand{\be}{\begin{equation}}
\newcommand{\ee}{\end{equation}}

\newcommand{\Rmnum}[1]{\expandafter\@slowromancap\romannumeral #1@}
\newcommand{\bea}{\begin{eqnarray}}
\newcommand{\eea}{\end{eqnarray}}
\begin{document}
\def\C{{\mathbb{C}}}
\def\R{{\mathbb{R}}}
\def\s{{\mathbb{S}}}
\def\T{{\mathbb{T}}}
\def\Z{{\mathbb{Z}}}
\def\W{{\mathbb{W}}}
\def\Bbb{\mathbb}
\def\BZ{\Bbb Z} \def\BR{\Bbb R}
\def\BW{\Bbb W}
\def\BM{\Bbb M}
\def\BC{\Bbb C} \def\BP{\Bbb P}
\def\CP{\BC\BP}
\begin{titlepage}
\title{Small Anisotropy in Stellar Objects in Modified Theories of Gravity} \author{} 
\date{
Shaswata Chowdhury, Tapobrata Sarkar 
\thanks{\noindent 
E-mail:~ shaswata, tapo @iitk.ac.in} 
\vskip0.4cm 
{\sl Department of Physics, \\ 
Indian Institute of Technology,\\ 
Kanpur 208016, \\ 
India}} 
\maketitle

\abstract{
Interior structures of stellar objects might have small pressure anisotropy due to several reasons, including rotation and the presence
of magnetic fields. Here, retaining the approximation of spherical symmetry, we study the possible role of small anisotropy in stellar 
interiors in theories of modified gravity, that are known to alter the hydrostatic equilibrium condition inside stars. 
We show how anisotropy may put lower and upper bounds on the modified gravity parameter depending on the polytropic equation
of state, and determine them numerically. 
We also study the mass of stellar objects in these theories, assuming such
equations of state, and find that the Chandrasekhar mass
limit in white dwarf stars gets substantially modified compared to the isotropic case, even without assuming 
the presence of extreme magnetic fields. Effects of small pressure anisotropy on the Hydrogen burning limit in low mass
stars are also briefly commented upon. It is shown that here the isotropic case can predict a theoretical lower bound on the scalar 
tensor parameter, in addition to a known upper bound.}
\end{titlepage}

\section{Introduction}

Einstein's general theory of relativity (GR) is one of the most successful theories of gravity till date, and its low energy
Newtonian limit is extensively used in the study of stellar structure and dynamics. Over the
last decades, it has been realised however that explaining issues relating to dark energy and dark matter possibly require
extending GR beyond the conventional starting point, i.e. the Einstein-Hilbert action. 
Indeed, the observed cosmic accelaration might indicate a shortcoming of GR, and 
point towards the necessity of such modified theories of gravity.
Several such theories are popular in the literature, one of the most important ones being scalar-tensor (ST) theories of gravity, that
arise by incorporating scalar fields in the Einstein-Hilbert action. 
In such theories, Newton's ``constant'' is no longer a constant, but a function of scalar fields appearing in the theory. 
In these situations, the corresponding Newtonian limit of GR also gets modified, and its applications to stellar dynamics
need to be carefully analysed, as these put observational bounds on the parameters of ST theories. Such studies have
been initiated in the literature fairly recently. 

Now, it is well known that stellar objects may not be isotropic -- a subject that has been discussed over decades. Anisotropy 
is a possible outcome in GR, the most famous example being the Einstein cluster geometries that
are sustained purely by tangential stresses. In the Newtonian context, anisotropy in stellar structures 
might exist due to several reasons, including rotation and the presence of magnetic fields in the stellar interior, which
might tend to oblate (or prolate) the stellar structure. Observations indicate that such distortations might be small, 
but nonetheless they form an integral part of the study of stellar dynamics. An important and interesting question in this 
context is then the role of (small) anisotropy in modified theories of gravity in the Newtonian limit.

The purpose of this paper is to revisit the problem of the modification of gravity inside stellar objects, via a 
model that includes small pressure anistropy in stellar matter. 
In the absence of a precise parametrization of such effects in stellar interiors, considering a fully general situation might 
be complicated. Useful information can nevertheless be gleaned in simplified scenarios, 
by considering simple but physically motivated models of such anisotropy and treating
them in conjuction with small parameters in the theory. This is what we do in this paper. In particular, we will retain the
assumption of spherical symmetry and then use the modified hydrostatic equilibrium condition along with small
anisotropy to models of stellar interiors via appropriate equations of state. 

This is done here for three different equations of state corresponding to a) degenerate non-relativistic electrons as appropriate
for the stellar core of brown and M-dwarf stars, b) degenerate relativisitic electrons that naturally appear in white dwarf 
scenarios allowing for moderately strong magnetic fields and c) degenerate relativistic electrons in large magnetic fields
corresponding to strongly magnetized white dwarfs that has been of interest of late. 
We first study the problem in general, and show how the natural lower cut-off for the ST parameter reported earlier in the literature is modified in the 
presence of anisotropy. Next, we solve the Lane-Emden equations numerically for different equations of state and show that there is also an upper
bound for that parameter for equations of state relevant to white dwarf stars. Then, within the small anisotropy regime, we 
compute stellar masses using appropriate
polytropic equations of state. It is shown that Chandrasekhar limit gets substantially modified in ST theories, in 
conjunction with anisotropy, without extreme magnetic fields. Finally, we make a few comments on the hydrogen burning
mass limit in low mass stars in these theories, and show that it gives a theoretical lower bound on the ST parameter in 
the isotropic case.

\section{Scalar Tensor Theories and Anisotropy in Stellar Mass Objects}

We first review some known facts about ST theories. 
One of the main consequences of such theories is the modification of gravitational forces inside stellar objects,
that is reflected in a change in the hydrostatic equilibrium condition. In the Newtonian limit, with $P$ being
the pressure, $\rho$ the density, $M$ the mass up to radius $r$ and $G$ the Newton's constant, this is given as 
\begin{equation}
\frac{dP}{dr} = - \frac{GM\rho}{r^2} - \Upsilon\left(\frac{G\rho}{4}\right)\frac{d^2M}{dr^2}~.
\label{TOV}
\end{equation}
Here, $\Upsilon$ is a dimensionless parameter that arises from the particular ST theory being considered, and is 
indicative of the deviation from GR, with $\Upsilon = 0$ denoting the Newtonian limit of conventional GR. 
$\Upsilon$ is related to the parameters that appear in an effective field theory of dark energy (see \cite{SaksteinPRD} for details). 
A possible measure
or a constraint on $\Upsilon$ from observed data can thus give significant information on the nature of ST theories 
that are admissible in nature. For example, near the center of a star, if we approximate $M(r) \sim 4\pi\rho_c r^3/3$ with
$\rho_c$ being the central density, then with $d^2M/dr^2 \sim 6M/r^2$, then we get
\begin{equation}
\frac{dP}{dr} = - \frac{M\rho_c}{r^2}G\left(1+\frac{3\Upsilon}{2}\right)~.
\label{TOVcore}
\end{equation}
Therefore, in order to retain the equilibrium conditions, we should have $\Upsilon > -2/3$ \cite{Saito}. On the other hand, if we
consider a typical low mass star, and assume that the surface gravity ($GM/r^2$) is constant to a good approximation, then,
close to the stellar radius we obtain $d^2M/dr^2 \sim 2M/r^2$. From eq.(\ref{TOV}), one now obtains a weaker constrain 
$\Upsilon > -2$. 

In any case, assuming that the parameter $\Upsilon$ is a constant throughout a stellar object, it is seen
that its effect is to redefine the Newton's constant \cite{SaksteinPRD}. 
Note that from eq.(\ref{TOVcore}), one can define a modified (renormalised) Newton's constant ${\tilde G} = G(1+3\Upsilon/2)$,
which implies that $G$ decreases with increasing $\Upsilon$, for a given ${\tilde G}$. However, in general, using
$dM/dr = 4\pi r^2\rho$ and writing eq.(\ref{TOV}) as
\begin{equation}
\frac{dP}{dr} = - \frac{GM\rho}{r^2} - \pi\Upsilon G\rho r\left(2\rho + r\frac{d\rho}{dr}\right)~,~
\label{Force}
\end{equation}
we see that the sign of the second term in eq.(\ref{Force}) will depend on the interplay between the two terms within the
bracket there. When $d\rho/dr$ is large and negative (away from the core), the second term might overtake the first,
and make the term involving $\Upsilon$ positive overall. This has also been noted in \cite{Saito}, and will depend
on the equation of state, as we will see. 

Recent work on the allowed values of $\Upsilon$ has been reported by Sakstein in \cite{Sakstein}, which considered the modification of 
Newtonian gravity inside low mass (brown dwarf) stars which are at the threshold of hydrogen burning in the core. 
Using the modified hydrostatic equilibrium condition of eq.(\ref{TOV}),
Sakstein obtained an expression for the minimum mass for hydrogen burning (MMHB) inside such stars, using the 
Lane-Emden equation. The end result for MMHB is then a function of $\Upsilon$, and appealing to known facts on MMHBs, 
Sakstein was able to put a bound on this parameter, which ruled out several classes of ST theories. Specifically, the
bound reads $\Upsilon \leq 1.6$. Later, Jain et. al. \cite{Jain} carried out an analysis of ST theories in white dwarf stars,
and came up with the more stringent bound $-0.18 < \Upsilon < 0.27$. 

We note here that the result of \cite{Jain}
is based on the mass-radius relations for white dwarf stars catalogued in \cite{Holberg}. On the other hand, \cite{Scalzo}
reported observations of supernovae of high luminosities, which are indicative of super-Chandrasekhar white dwarf
stars of masses above $2M_{\odot}$, significantly higher than the Chandrasekhar limit of $\sim 1.44 M_{\odot}$. Such high
mass progenitors are thought to have strong magnetic fields in their interiors, and in part motivate the study of anisotropy. 

Indeed, if in eq.(\ref{TOV}), one allows for an additional pressure anisotropy term on the right hand side with the assumption that
spherical symmetry is still maintained, the situation might change qualitatively. Such situations, which arise when the 
radial and tangential pressures are not equal inside a non-rotating stellar body, are well known in GR \cite{MakHarko}. 
It is also well known that these can arise out of gravitational collapse scenarios. In a Newtonain formalism in the presence of 
anisotropy in a spherically symmetric situation, where $P_r \neq P_{\theta} = P_{\phi}$, eq.(\ref{TOV}) is modified to a more general 
form including an anisotropy term $\Delta(r)$ \cite{BowersLiang}
\begin{equation}
\frac{dP_r}{dr} = - \frac{GM\rho}{r^2} - \Upsilon\left(\frac{G\rho}{4}\right)\frac{d^2M}{dr^2} + \Delta(r)~,~~
\Delta(r)= \frac{2}{r}\left(P_{\perp}-P_{r}\right)~.
\label{TOVA}
\end{equation}
where we have denoted the tangential pressure as $P_{\perp}$. In terms of Cartesian coordinates, this implies that 
at each point inside the stellar object, we rotate the axes, such that the pressures along two of the Cartesian 
directions are the same, and that these are not equal to the pressure in the third direction. 

As mentioned before, 
anisotropy can occur in stellar objects due to several reasons. Let us briefly recall some of them, following \cite{HerreraSantosReview}. 
For the situations that we will be interested in this paper, the
important ones include : 1) the stellar fluid being made up of two (or more) isotropic fluids whose mixture might nonetheless be 
anisotropic. Such two-fluid models have been extensively investigated in astrophysics and 
cosmology, since the early works of \cite{Letelier} and \cite{Bayin}. 2) The stellar object might have (slow) rotation. Rotating
polytropes have been studied since the works of \cite{ChandraRot} (see \cite{KippenWei} for a detailed account). Rotation
induces anisotropy via a centrifugal force that is proportional to the radial distance $r$, and is formally identical to 
$P_{\perp}-P_r \propto r^2$. 3) The presence of a magnetic field
inside the star. This issue has been intensely debated over the last few years, after the observation of super-Chandrasekhar
progenitor white dwarfs, in the mass range of $2.1-2.8~M_{\odot}$, see \cite{Scalzo}, \cite{Hachisu} and references therein. 

We recall that \cite{Bani1} proposed a strongly magnetized white dwarf scenario with 
central magnetic fields $\sim 10^{16}~{\rm G}$ as a possible candidate for such stars (for criticisms
of their model, including possible stability issues, see \cite{Konar}). We also point out that on the other hand, 
a theoretical bound on the magnetic
field inside low mass stars (for mass $M \sim 0.3 M_{\odot}$) reads $B_{max} \sim 10^6~{\rm G}$ \cite{Browning}. 
Very recently, evidence for
magnetic activity (magnetic fields $\sim 10^3-10^4~{\rm G}$ at the surface) in low density brown dwarfs has been reported as well.
in \cite{BDMag}. These points (1) -- (3) above constitute the motivation for studying anisotropy in stellar objects in this work. 

Anisotropic effects inside stellar matter due to constant magnetic fields is an extremely well studied subject, starting from a series of
papers by Canuto and Chiu \cite{Canuto1},\cite{Canuto2},\cite{Canuto3}. While this was done in the framework of second
quantization, Ferrer et. al. \cite{Ferrer} performed the analysis in the language of quantum mechanics which makes the results somewhat
more transparent. Let us briefly recall their results which will
be important for us. The authors of \cite{Canuto1} worked out the general theory of an electron gas by solving the Dirac equation 
in a strong magnetic field and obtained the equation of state, and further showed that the energy-momentum tensor
is anisotropic. The Dirac equation has to be used, in contrast to the Schrodinger equation with a spin-magnetic field coupling
introduced by hand, when the cyclotron frequency becomes comparable to  to the rest energy of the electron, 
i.e $e\hbar B_c/m_e c \sim m_e c^2$, with $B$ being the applied magnetic
field and $m_e$ is the electron mass. This relation implies that $B_c \simeq 4 \times 10^{13}~{\rm G}$. In terms of this critical
magnetic field,  \cite{Canuto1} showed that for both relativistic and non-relativistic electrons, in the degenerate limit, 
quantum effects set in when 
\begin{equation}
\left(\frac{\rho}{\mu_e \times 10^7~{\rm gm~cm^{-3}}}\right) \leq 2\frac{B}{B_c}~.
\label{qlimit}
\end{equation}
For brown dwarf stars, taking typical core densities as $\rho_c \sim 10^2~{\rm gm~cm^{-3}}$, one obtains 
$B \sim 10^9 - 10^{10}~{\rm G}$ for quantum effects to set in. This is of course way beyond the surface magnetic fields
reported till now. On the other hand, for white dwarf stars, quantum effects become prominet for $B \sim 10^{13} - 10^{14}~{\rm G}$,
as is obtained by typical estimates of $\rho_c \sim 10^8~{\rm gm~cm^{-3}}$.

On the other hand, thermodymaics of degenerate electron gases in the presence of 
high magnetic fields typical to white dwarf stars was performed in \cite{Canuto2}, where it was shown that under sufficiently strong magnetic
fields, when only the first Landau level is occupied, the pressures tangential to the magnetic field direction might vanish. 
although it was argued that in general, effects of temperature might prevent a possible collapse due to this condition. 
The analysis of \cite{Ferrer} on the other hand shows that 
\begin{equation}
P_{\perp}-P_{\parallel} = -B {\mathcal M} + \frac{B^2}{4\pi}~,
\label{Chaichian1}
\end{equation}
where $P_{\parallel}$ is the pressure along the direction of the magnetic field, $P_{\perp}$ is the pressure in the
other directions, $B$ is the applied field and ${\mathcal M}$ is the magnetization of matter, that can be computed from
the thermodynamic potential (grand partition function). In this context, \cite{Canuto3} computed
the magnetic moment of magnetised Fermi gases and found that for a given magnetic field $B$, 
the maximum value ${\mathcal M}_{max}$ of the magnetization satisfies
\begin{equation}
{\mathcal M}_{max} \sim 10^{-3}B \implies \left(B{\mathcal M}\right)_{max} \sim 10^{-3}B^2~.
\end{equation}
The term involving magnetization is in any case small (and known to be oscillatory) and will be subdominant. 
We therefore have that in case of constant classical magnetic fields, one can have a pressure anisotropy of $B^2/(4\pi)$. 

Of course, the analysis of Canuto and Chiu as well as Ferrer et. al. 
were for constant magnetic fields which break spherical symmetry, and should
be incorporated in stellar dynamics in terms of 
cylindrical (or more appropriately oblate-spheroidal) coordinates. Such a route might be substantially more 
complicated than a simple minded one in which one retains spherical symmetry as an approximation and models the 
effect of anisotropy coming from all possible sources listed before, by an inherent pressure 
asymmetry of the form given in eq.(\ref{TOVA}), while
being mindful of the estimates of anisotropy as follows from our discussion above. It is this approach that we will follow. 

In order to proceed, one has to choose a specific form for $\Delta(r)$. Two of the popular models considered
in the literature are due to Heintzmann and Hillebrandt \cite{Heintzmann} who model the anisotropy as 
$P_{\perp} - P_r \propto P_r(r)$ and Herrera and Santos \cite{HerreraSantos}, \cite{HerreraSantosReview}
who use a polynomial of the form $P_{\perp} - P_r \propto r^n$. The latter model has been studied extensively (for a
recent work see \cite{HerreraBarreto}). In this model, in order for the anisotropy to vanish at the origin
which is known to be a requirement in theories with non-vanishing core density and pressure \cite{Madsen}, we require
$n>1$,  in which case the anisotropy increases as a power law of $r$ as one moves away from the center. 
This is thus more suited to modelling rotational effects only. 

On the other hand, 
the model of Heintzmann and Hillebrandt offers a couple of advantages. First, we can control the dimensionless $\beta(r)$ 
to define the anisotropy. Second, with $\beta(r)$ being zero at the core, the anisotropy rises with $r$, but tapers off
to zero at the surface of the star where the radial pressure is zero. If we think of the anisotropy in our modified gravity
models as originating at least partly from the magnetic field, this is a better model -- indeed the magnetic fields at the surface of 
stars are believed to be orders of magnitude lower than those at the core, so we can effectively treat it as small compared 
to the anisotropy near the core. Thus, within the approximation of spherical symmetry, we consider a model in which the 
net anisotropy (due to a possible combination of all reasons mentioned before) is proportional to the local radial pressure, 
and differs from the core pressure $P_c$ by a small amount, i.e we will restrict ourselves to situations where
$P_{\perp} - P_r \ll P_c$. 
This is then our working model, following \cite{Heintzmann}.

\section{Anisotropy Effects in Scalar-Tensor Theories}

We will henceforth adopt, as a simple model, 
$P_{\perp}-P_{r} = \beta(r) P_r$, where $\beta$ is a dimensionless measure of the strength of the anisotropy, and 
we allow the freedom of $\beta$ being a function of the radial distance, this being a non-trivial modification of 
the form of anisotropy originally used by \cite{Heintzmann} and will be crucial for us, as we will see. 
Physicality of the situation, i.e the derivative of $P_r$ vanishes in the limit $r \to 0$, 
demands that in this limit, $(P_{\perp}-P_{r})/r \to 0$, which
in turn implies that $\beta(r)P_r \to 0$ at the center faster than $r$. 
That this must be the case was recognized in the analysis of \cite{Madsen}. For a reasonable polytropic equation of state, 
this should also hold true for the density near the center, and we will thus assume that the derivative of the density also
vanishes in the same limit. Expanded around the center, the $r$-dependence of 
the radial pressure and density profiles are thus of ${\mathcal O}(r^2)$
and higher. We will also assume a polytropic equation of state for the matter inside \cite{Shapiro}, given as
\begin{equation}
P_r = K \rho^{\Gamma}~,~~\Gamma = \frac{n+1}{n}~.
\label{poly}
\end{equation}
The polytropic index $\Gamma$ will be left general as of now, and we will comment about it in a while. 
As is standard, we multiply eq.(\ref{TOVA}) by $r^2/\rho$, and use $dM/dr = 4\pi r^2\rho$, to obtain 
\begin{equation}
\frac{d}{dr}\left(\frac{r^2}{\rho}\frac{dP_r}{dr}\right) + 4\pi G r^2\left(\rho + \frac{3\Upsilon}{2}\rho + \frac{3\Upsilon }{2}r\frac{d\rho}{dr}
+ \frac{\Upsilon}{4}r^2\frac{d^2\rho}{dr^2}\right) - \frac{d}{dr}\left(\frac{2r}{\rho}\beta(r) P_r\right) = 0~.
\label{TOVB}
\end{equation}
We will treat $\Upsilon$ to be a constant. Next, we Taylor expand $P_r$, $\rho$, $\beta$ in terms of the 
variable $r/r_c$, with $r_c$ being a typical length scale, near $r=0$ and equate
terms of similar powers in $r$. We thus write 
\begin{equation}
\rho = \rho_{c} + \frac{1}{2}\rho_{2}\left(\frac{r}{r_c}\right)^2 + \cdots~,~~P_r = K\rho^{\Gamma}~,~~
\beta(r) = \beta_c + \beta_1\left(\frac{r}{r_c}\right) + \frac{1}{2}\beta_2\left(\frac{r}{r_c}\right)^2 + \cdots~,
\end{equation}
where$P_c$, $\rho_c$ and $\beta_c$ are central values of the pressure, density and anisotropy, with $P_c$ and $\rho_c$ 
being related by the equation of state, and we have
not made any assumption on the derivative of $\beta(r)$ near the core. 
In hindsight, we now make a specific choice of the length scale, 
\begin{equation}
r_c^2 = \frac{(n+1)P_c}{4\pi G\rho_c^2} \equiv \frac{K(n+1)\rho_c^{\Gamma - 2}}{4\pi G}~.
\label{rc}
\end{equation}
With the above ingredients, it can be seen that the ${\mathcal O}(r^0)$ term gives 
\begin{equation}
K\beta_0\rho_c^{\Gamma - 1}= 0~.
\end{equation}
Since the central density $\rho_c \neq 0$, this implies that the anisotropy at the center should vanish, i.e $\beta_0=0$ \cite{Madsen}. 
Next, the ${\mathcal O}(r)$ term is easily shown to imply that (with $\rho_{c}\neq 0$), $\beta_1 = 0$. 
With this input, the ${\mathcal O}(r^2)$ term then gives
\begin{equation}
2 \pi  G r_c^2 \left(2+3\Upsilon\right)\rho_c^3-3K\rho_c^{\Gamma}\left(2\beta_2\rho_c - \Gamma\rho_2\right)=0~~\to~~\beta_2 = 
\frac{n+1}{6}\left(2+3\Upsilon\right) + \Gamma\frac{\rho_2}{\rho_c}~.
\label{beta}
\end{equation}
With $\rho_{c}$ being the central radial pressure we must have $\rho_{2} < 0$, as the density should decrease away from the 
core. When $\beta = 0$, this implies that $\Upsilon > -2/3$ \cite{Saito}. Moreover, when $\Upsilon = 0$ but
we have non-zero anisotropy, this means that $\beta_2$ is bounded, i.e 
$\beta_2 < \frac{n+1}{3}$. 
Modified gravity however changes the scenario. Here, we find the constraint 
\begin{equation}
\Upsilon > -\frac{2}{3} + \frac{2\beta_2}{n+1}~.
\label{upscon1}
\end{equation}

We note that other models for $\beta(r)$ can also be considered. 
For example, one can consider models of anisotropy for which $\beta(r) = {\hat\beta}r^2(r_c-r)^2/r_c^4$.
In the first case, the pressure anisotropy $(P_{\perp} - P_r)/r$ ($\equiv \beta(r)P(r)/r$) starts from zero, reaches a 
maximum inside the stellar body and then tapers off to zero, since the pressure reduces as one approaches the
surface of the star and vanishes at the surface. In the second model, the pressure anisotropy vanishes at
$r=r_c$, then increases and vanishes again at the surface of the star. This is a more exotic model, and since the physical situation 
in which this can be justified is not quite clear, we will not analyse this further. 

Also, as a curiosity, we note that it is possible in this scenario to sustain gravitational stability entirely by anisotropic forces. This
is theoretically possible if the first two terms on the right hand side of eq.(\ref{TOVA}) combine to zero, i.e the 
mass satisfies the Euler's equation $\Upsilon d^2M/dr^2 = -4M/r^2$. In this case, as $r\to 0$, one can use $\Upsilon = -2/3$
and then integrate the pressure equation
directly to obtain $P = P_c~{\rm exp}(ar^2/r_c^2)$ with $a = \Gamma\rho_2/(2\rho_c)$ being negative, and further obtain the density 
from the equation of state. 
\begin{figure}[h!]
\hspace{0.3cm}
\begin{minipage}[b]{0.45\linewidth}
\centering
\centerline{\includegraphics[scale=0.3]{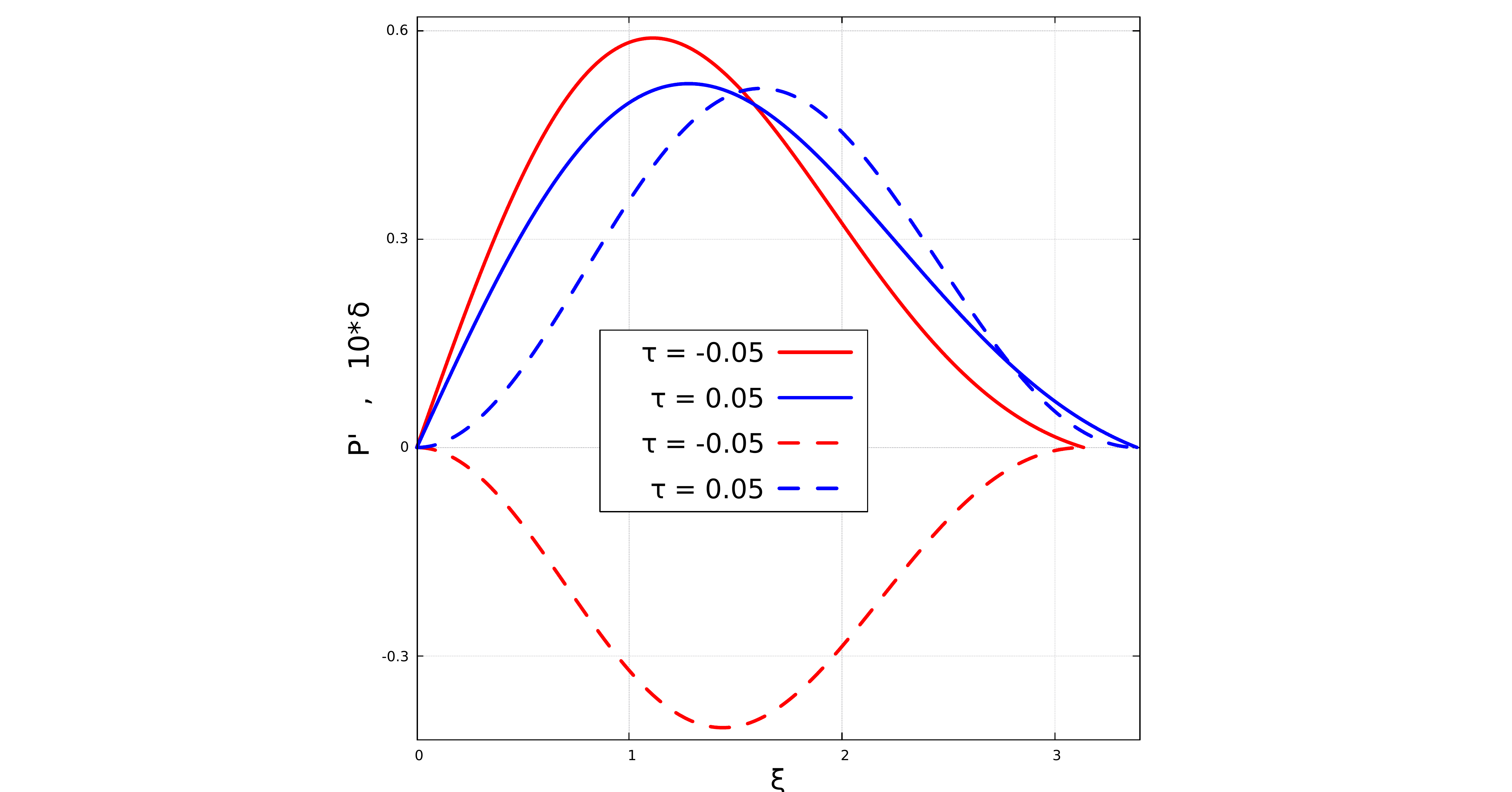}}
\caption{$|dP_r/dr|$ (solid lines) and $\delta$ magnified by $10$ (dashed lines) for $\Upsilon = 0.1, \Gamma = 2$.}
\label{dpdr1}
\end{minipage}
\hspace{0.5cm}
\begin{minipage}[b]{0.45\linewidth}
\centering
\centerline{\includegraphics[scale=0.3]{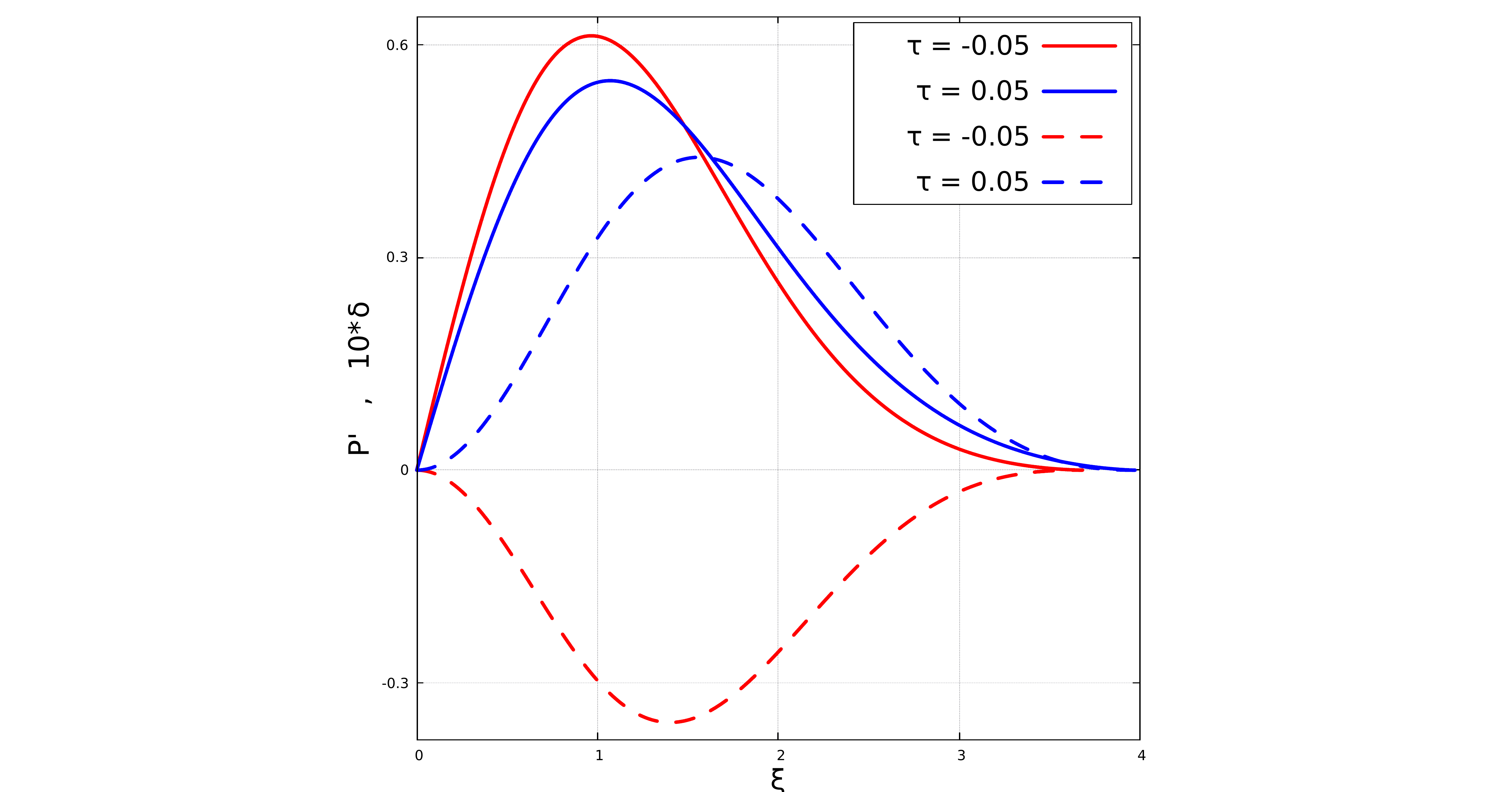}}
\caption{$|dP_r/dr|$ (solid lines) and $\delta$ magnified by $10$ (dashed lines) for $\Upsilon = 0.1, \Gamma = 5/3$.}
\label{dpdr3by2}
\end{minipage}
\end{figure}

To connect the above discussion with relevant stellar physics, we will now set up the dimensionless Lane-Emden equation for our system. 
This is straightforward using eq.(\ref{TOVA}), once we assume a polytropic equation of 
state (eq.(\ref{poly})) and introduce dimensionless variables
$\xi$ and $\theta$, such that 
\begin{equation}
\rho = \rho_c \theta^n~,~~ r = \xi r_c~,~~P_r = K\rho_c^{\Gamma}\theta^{n+1}~,
\label{rhop}
\end{equation}
with $r_c$ given by eq.(\ref{rc}). Then, the Lane-Emden equation takes the form
\begin{equation}
\frac{1}{\xi^2}\frac{d}{d\xi}\left[\xi^2\frac{d\theta}{d\xi} - \frac{2}{n+1}\tau\xi^3\theta + \frac{\Upsilon}{4}\left(2\xi^3\theta^n + n\xi^4
\theta^{n-1}\frac{d\theta}{d\xi} \right)\right] + \theta^n=0~,~~
\label{LEE}
\end{equation}
where we have defined the dimensionless paramter $\tau = (1/2)\beta_2$. This equation has to be solved numerically
along with the boundary conditions $\theta(0)=1$ and $\theta'(0)=0$. 

Note that from the discussion above, it follows that in our choice of anisotropy,
\begin{equation}
\frac{P_{\perp} - P_r}{P_c} = \tau\xi^2\frac{P_r}{P_c}~.
\label{anisosmall}
\end{equation}
Near the core with small $\xi$, the right hand side of eq.(\ref{anisosmall}) is small. As $\xi$ increases, the pressure
decreases away from the core, and when $\xi$ reaches its boundary (largest) value, the pressure has almost dropped to
zero so that this quantity is again small. In order to make the pressure anisotropy small throughout, we will henceforth
choose $|\tau| \lesssim 10^{-2}$, so that the condition of small anisotropy $(P_{\perp} - P_r)/P_c \ll 1$ is always satisfied. 

In figures (\ref{dpdr1}) and (\ref{dpdr3by2}), we have plotted $P'=|dP_r/dr|$ in units of $K\rho_c^{\Gamma}/r_c$ (solid lines)
and the anisotropy $\delta(r) = (P_{\perp} - P_r)/P_c$ (dashed lines) with $\Upsilon = 0.1$ in both cases and 
$\Gamma=2$ and $5/3$, respectively. These quantities have been plotted as a function of the variable $\xi$, and $\delta$ 
has been multiplied by a factor of $10$ to offer better visibility. 
In both figures, the red line correspond to $\tau = -0.01$ and the blue ones to $\tau = 0.01$, respectively. We see that
the anisotropy peaks close to the maximum of the pressure gradient. Numerically, the maximum value of $\delta$ is seen
to be $\simeq 0.05$ for $\Gamma=2$, and $\simeq 0.04$ for $\Gamma=5/3$. We thus see that the anisotropy is indeed small,
as discussed after eq.(\ref{anisosmall}). We have checked that for other values of $\Upsilon$ also, similar estimates are
obtained.  
\begin{figure}[h!]
\hspace{0.3cm}
\begin{minipage}[b]{0.45\linewidth}
\centering
\centerline{\includegraphics[scale=0.3]{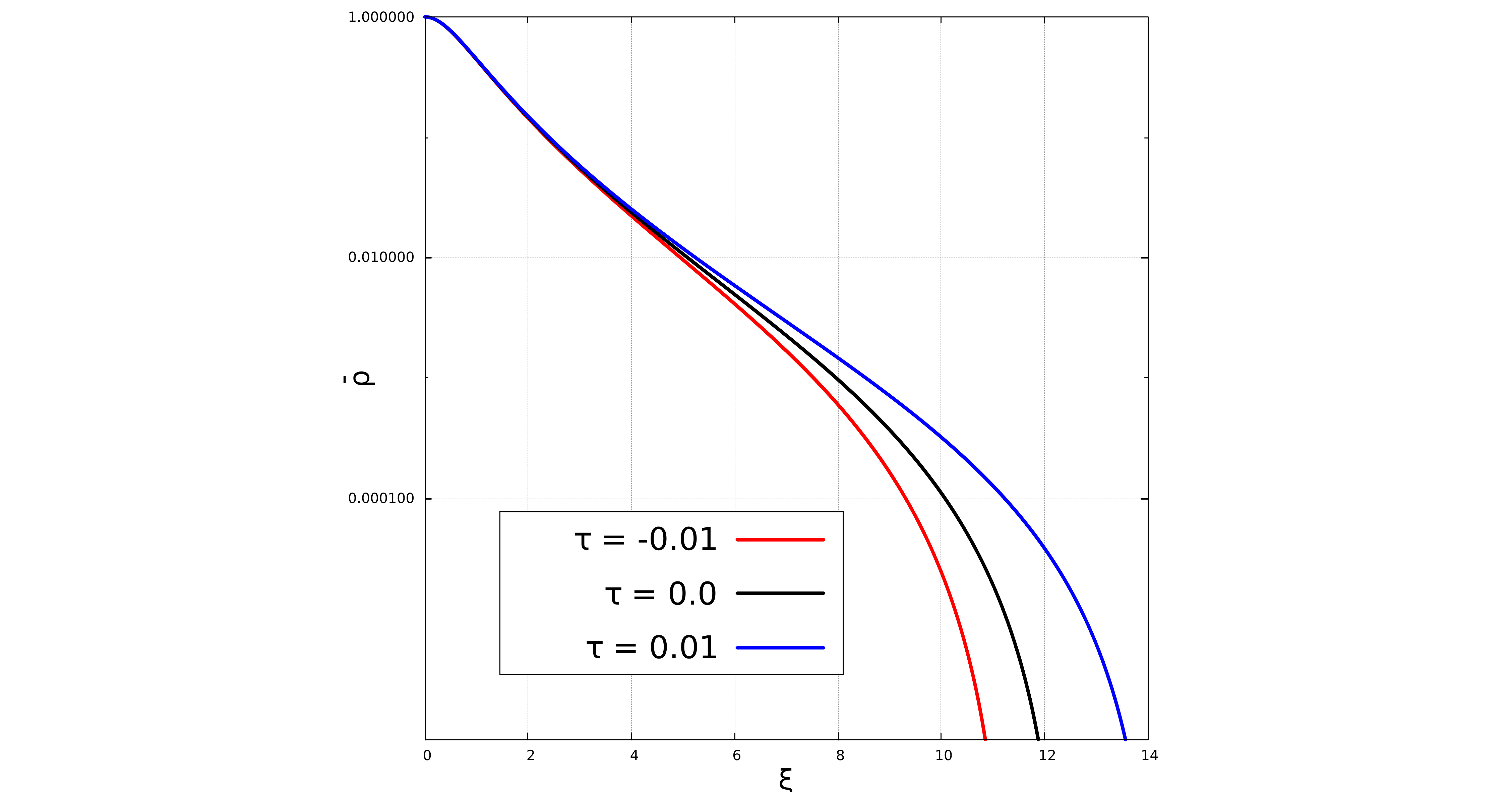}}
\caption{${\bar\rho}=\rho/\rho_c$ vs $\xi$ plots for $\Gamma = 4/3$ and $\Upsilon = 1$.}
\label{rhon3}
\end{minipage}
\hspace{0.5cm}
\begin{minipage}[b]{0.45\linewidth}
\centering
\centerline{\includegraphics[scale=0.3]{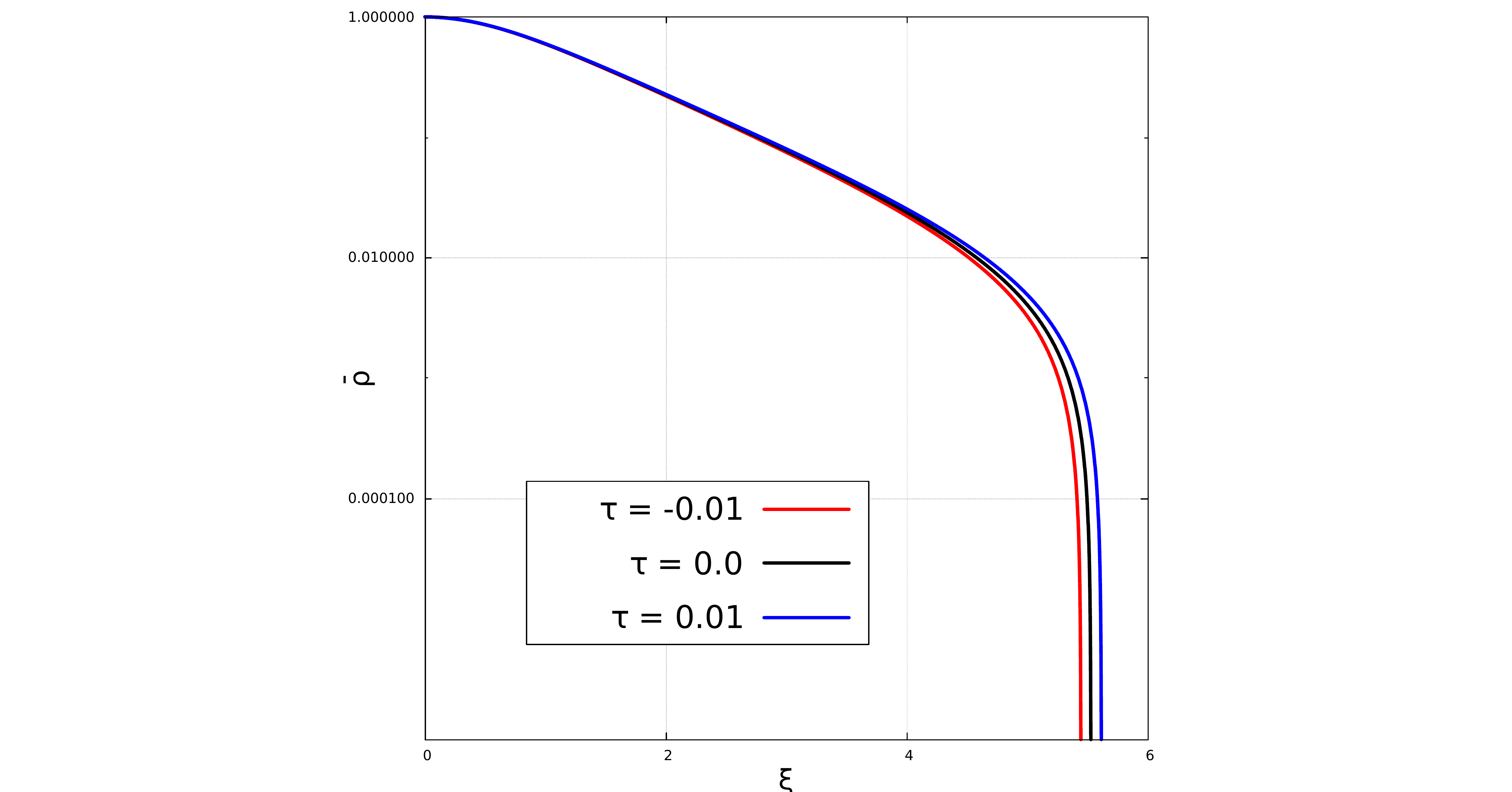}}
\caption{${\bar\rho}=\rho/\rho_c$ vs $\xi$ plots for $\Gamma = 5/3$ and $\Upsilon = 1$.}
\label{rhon3by2}
\end{minipage}
\end{figure}

Next, we show the results for the density for different values of $\Upsilon$ and $\tau$. In fig.(\ref{rhon3}),
we show ${\bar \rho} = \rho/\rho_c$ as a function of $\xi$ for $\Upsilon = 1$ and for different values of $\tau$ for $\Gamma = 3$. 
A similar plot is shown in fig.(\ref{rhon3by2}) where we have taken $\Gamma = 3/2$. 
The above plots illustrate what we have said before -- with $r d\rho/dr = \xi d\rho/d\xi$ behaving differently for different
equations of state, the second term on the right hand side of eq.(\ref{Force}) might change sign depending on the precise
solution of the Lane-Emden equation. 

Now we establish that solutions of the Lane-Emden equations imply an upper bound on $\Upsilon$ as a function of the
anisotropy parameter. This is due to the following reason. From numerical analysis of eq.(\ref{LEE}), we find that for a given 
value of $\tau$, $\theta(\xi)$ has a turning point beyond some particular value of $\Upsilon$, at a radial distance less than the radius of
the star. This would imply, given the
definitions of $\rho$ and $P$, that the density has a turning point (it actually has a second turning point and 
goes to zero at the boundary beyond that). On physical grounds, these cases have to be ruled out, as 
for any sensible density profile, the density should decrease continuously from the core up to the surface of the star, without
any pressure belts. We now give an explanation for this. From eq.(\ref{TOVA}), a turning point in the pressure (equivalently
the denity via the polytropic equation of state) is generically given by 
\begin{equation}
-\frac{GM\rho}{r^2} - 2\pi r G \Upsilon\rho^2 + \Delta(r) = 0~,
\end{equation}
where we have used the definition of the density, and also set $d\rho/dr = 0$. Using $\Delta = 2\beta(r)P_r/r$, eq.(\ref{rhop}), and 
the definition of $r_c$ given in eq.(\ref{rc}), it can be checked that the existence of a turning point implies that 
\begin{equation}
\Upsilon \lesssim -\frac{2}{3} + \frac{4\tau}{n+1} \theta_T^{1-n}~,
\label{upscon}
\end{equation}
where $\theta_T$ is the value of $\theta$ at the turning point.
With $\theta_T=1$ (which is the boundary condition at the origin), we recover back eq.(\ref{upscon1}) which
was derived close to the center, in order to {\it avoid} any point of inflection near the center. Note that eq.(\ref{upscon1}) was
derived assuming that the density decreases close to the origin. Here, on the other hand, the density should increase
if there is a local minimum, and this puts an upper bound on $\Upsilon$. 
We see that for $\tau \leq 0$, eq.(\ref{upscon}) clearly contradicts eq.(\ref{upscon1}), so that there are no
turning points of $\theta$ inside the stellar surface. 
\begin{figure}[h!]
\hspace{0.3cm}
\begin{minipage}[b]{0.45\linewidth}
\centering
\centerline{\includegraphics[scale=0.3]{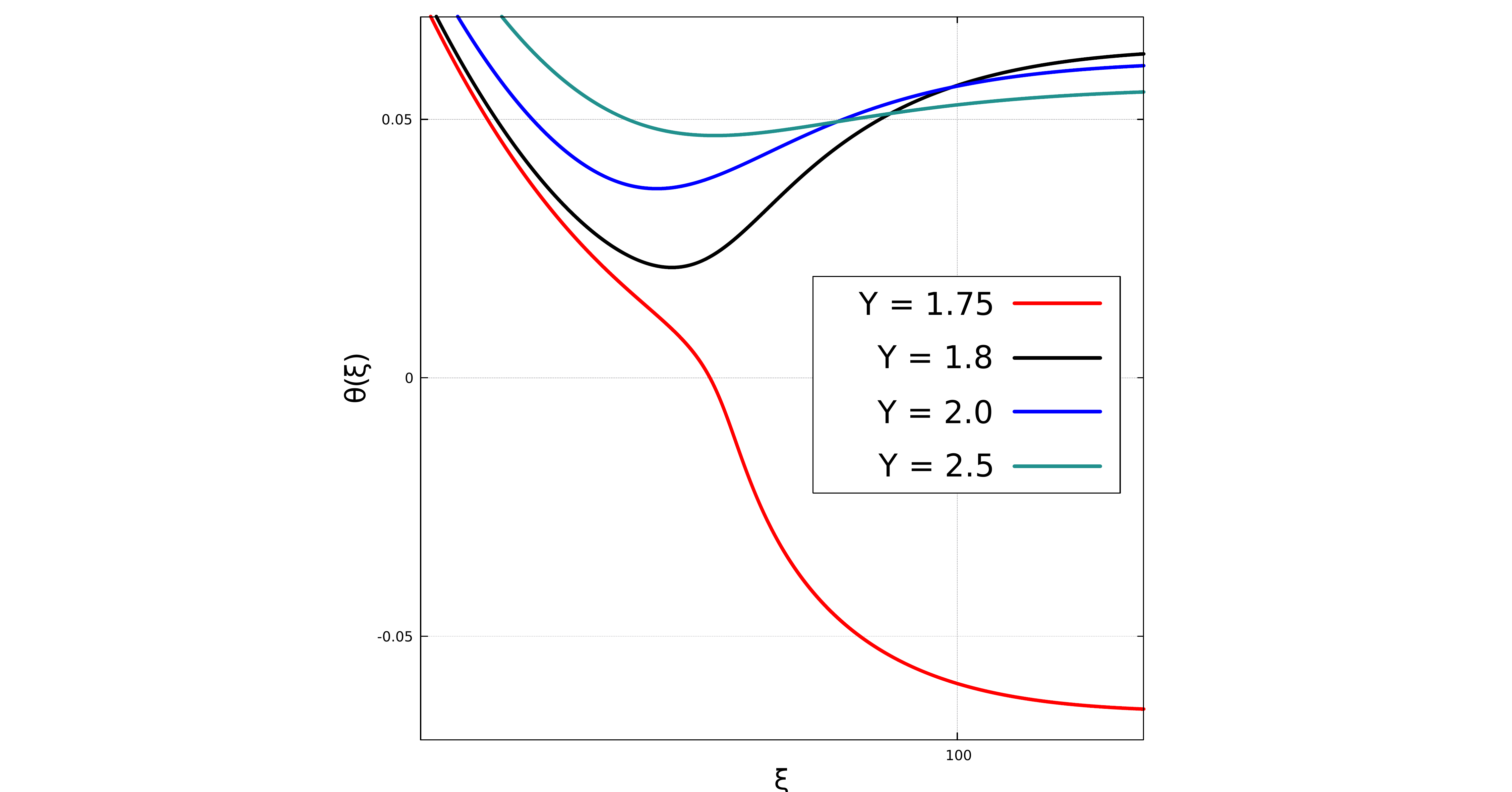}}
\caption{$\theta$ vs $\xi$ plots for $\Gamma = 4/3$ and $\tau = 0.01$.}
\label{turning}
\end{minipage}
\hspace{0.5cm}
\begin{minipage}[b]{0.45\linewidth}
\centering
\centerline{\includegraphics[scale=0.3]{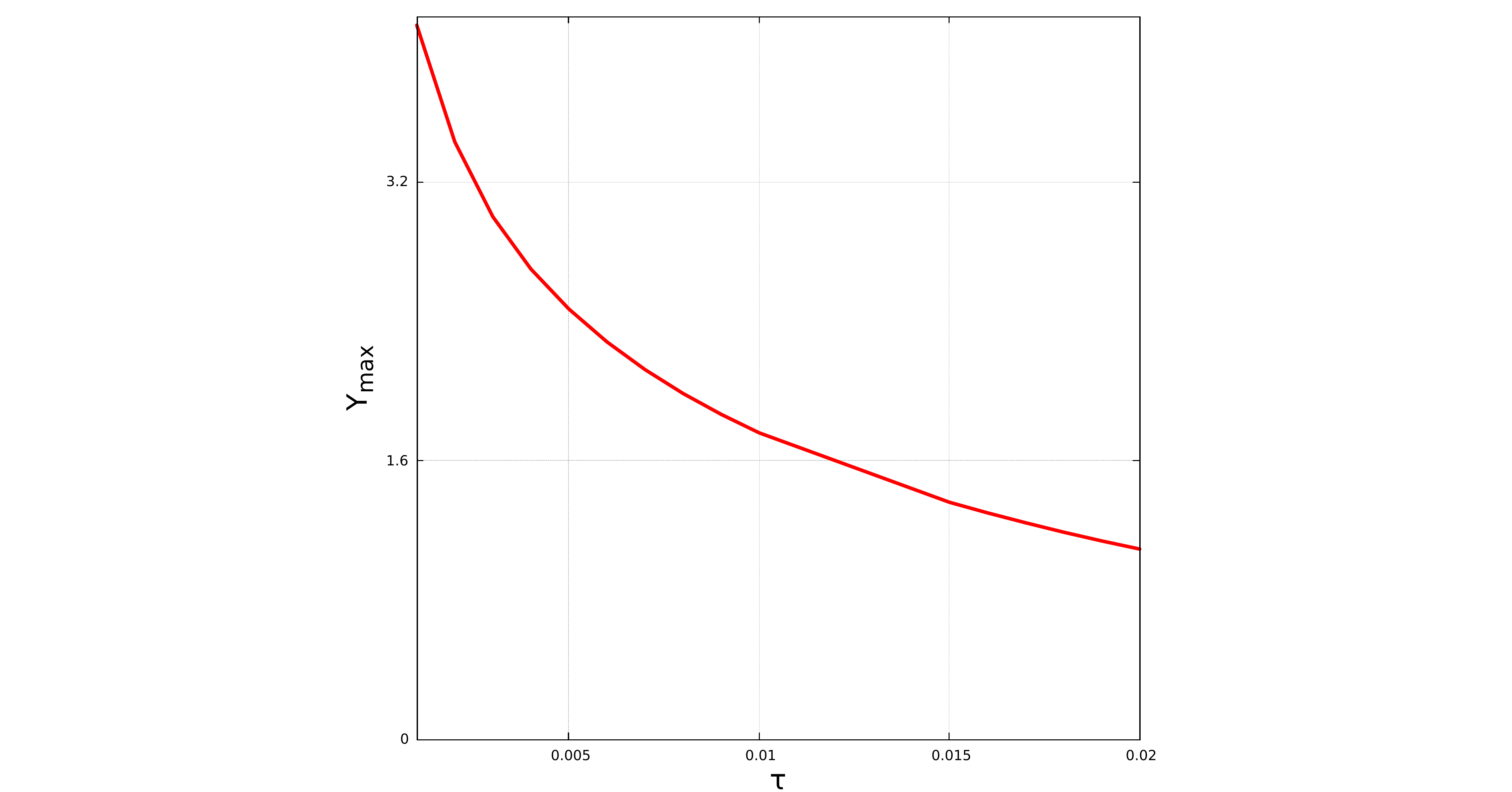}}
\caption{$\Upsilon_{max}$ vs $\tau$ plots for $\Gamma = 4/3$.}
\label{upsmax}
\end{minipage}
\end{figure}

For positive values of $\tau$, the situation becomes more interesting. 
Results of our numerical analysis for the Lane-Emden equation in this case 
are presented in fig.(\ref{turning}). Here, we have taken $\Gamma = 4/3$, $\tau = 0.01$,
and plotted $\theta$ as a function of $\xi$ for various values of $\Upsilon$. We see that beyond $\Upsilon \sim 1.75$, the
$\theta$ profile develops a turning point inside the stellar radius. This means that the density and the pressure will also
develop such a turning point, and so $(\tau = 0.01, \Upsilon \gtrsim 1.75)$ has to be
ruled out. We carried out this analysis for several values of $\tau$, and the results are depicted in fig.(\ref{upsmax}), where
we have plotted the maximum value of $\Upsilon$, admissible for a given value of $\tau$. It might seem that sufficiently large positive
values of $\tau$ would nullify the ST theory parameter, but such high anisotropy might be difficult to explain within the
spherically symmetric approximation that we are using. 

One might wonder if similar turning points exist inside the stellar radius for other polytropic equations of state. 
That this is not the case can be easily seen for $n=1$ ($\Gamma=2$), where the condition
in eq.(\ref{upscon}) becomes independent of $\theta_T$, which imples that there will be no turning point of $\theta$ inside the stellar
surface, given eq.(\ref{upscon1}). For $n=3/2$ ($\Gamma = 5/3$), our numerical analysis indicates that although theoretically
possible, such turning points inside the stellar radius 
exist only for values of $\Upsilon$ and $\tau$ which make the stellar radius unrealistically large, and must therefore 
be excluded from the analysis. 

\section{Anisotropy Effects on Stellar Mass}

We will now briefly analyse the variation of the mass of stellar objects as a function of the ST parameter $\Upsilon$ and the anisotropy
parameter $\tau$, using the mathematical formalism discussed in the previous section. We will restrict ourselves to 
white and brown dwarfs in this analysis. Note that the LEE determines the mass of the stellar object via
\begin{equation}
M = 4\pi\int_0^R \rho(r) r^2 dr = -4\pi r_c^3 \rho_c \xi_R^2\theta'(\xi_R)~~({\rm for}~ n \neq 1),~ = 
-4\pi r_c^3 \rho_c \xi_R^2\left(1 + \Upsilon\xi_R^2\right)\theta'(\xi_R)~~({\rm for} ~ n = 1)~,
\label{MCh}
\end{equation}
where the radius of the star is related to $\xi_R$ by $R=r_c\xi_R$. We will focus on the dimensionless quantity
$M/M_{Ch}$ where $M_{Ch}$ is the Chandrasekhar mass defined by the expressions in eq.(\ref{MCh}) with 
$\Upsilon = \tau = 0$. 
\begin{figure}[h!]
\hspace{0.3cm}
\begin{minipage}[b]{0.45\linewidth}
\centering
\centerline{\includegraphics[scale=0.3]{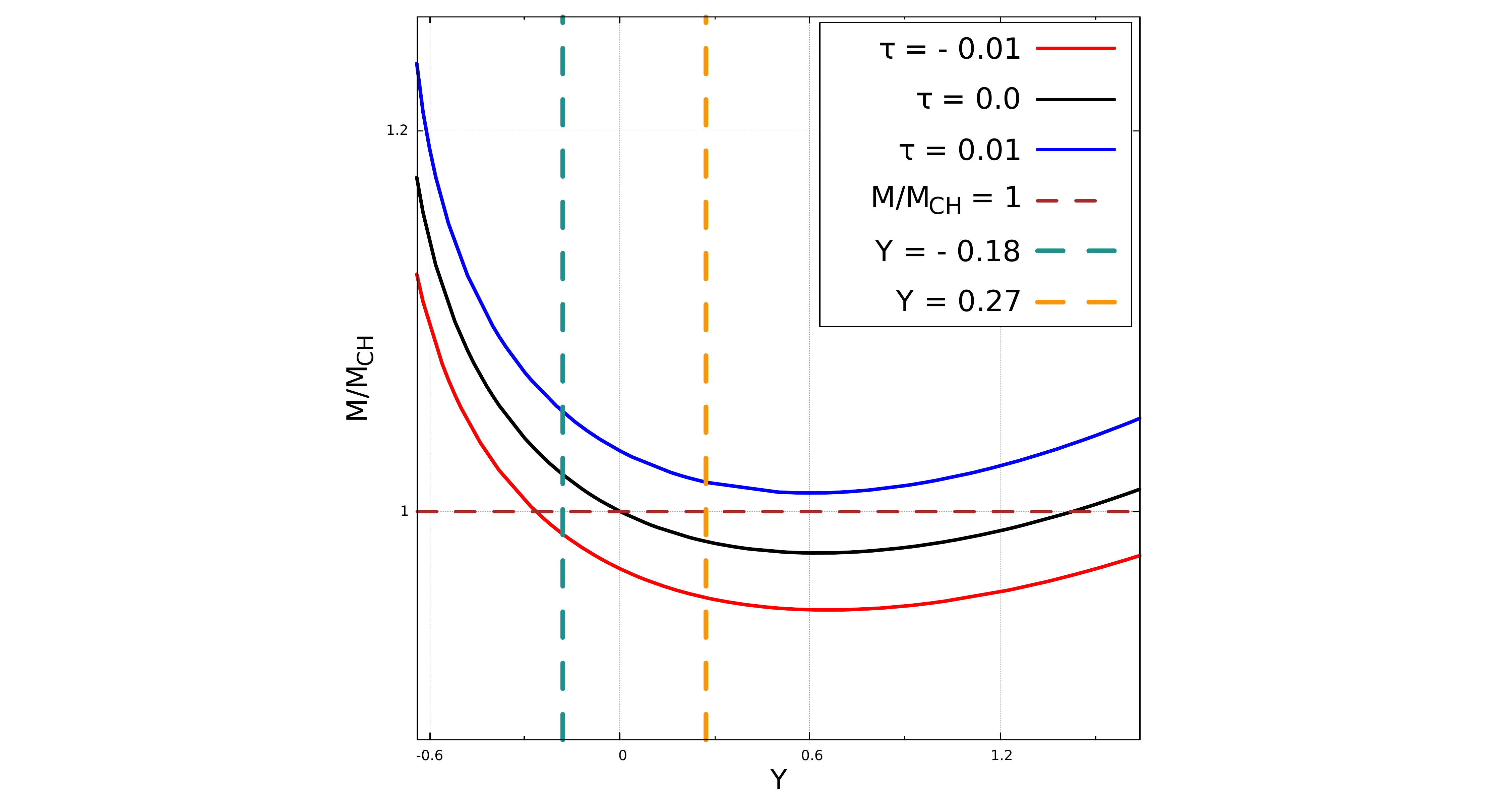}}
\caption{$M/M_{Ch}$ vs $\Upsilon$ for $\Gamma = 5/3$}
\label{mratio3by2}
\end{minipage}
\hspace{0.5cm}
\begin{minipage}[b]{0.45\linewidth}
\centering
\centerline{\includegraphics[scale=0.3]{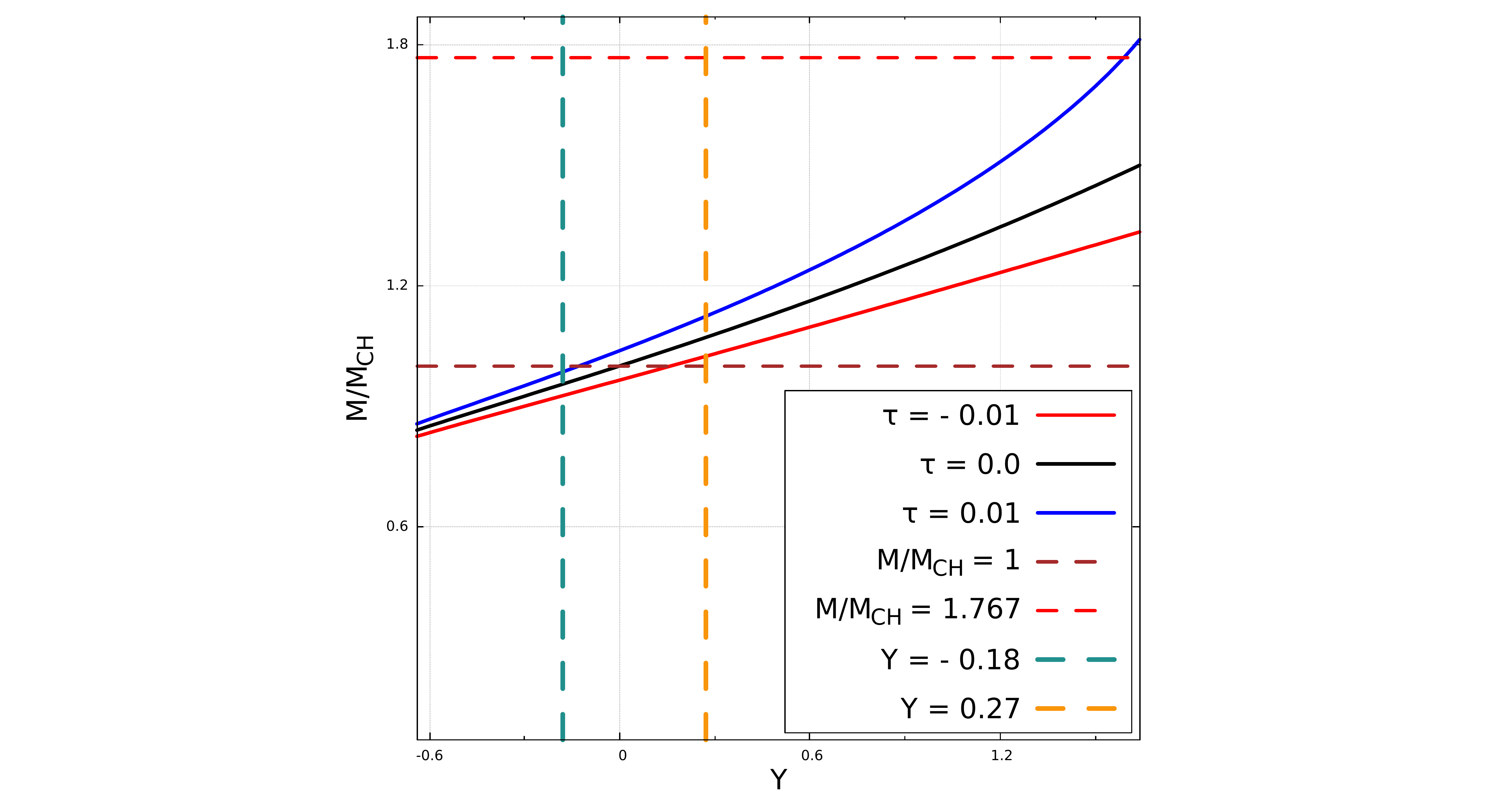}}
\caption{$M/M_{Ch}$ vs $\Upsilon$ for $\Gamma = 4/3$}
\label{mratio3}
\end{minipage}
\end{figure}
We compute this quantity for three cases, $\Gamma = 5/3,~4/3$ and $2$. In the next three graphs, for ready comparison
with existing literature, we indicate the current bounds on $\Upsilon$. The extreme left 
limit is indicative of the lower bound $\Upsilon = -2/3$, obtained in \cite{Saito}. The extreme right
limit is $\Upsilon = 1.6$ that arises from the analysis of \cite{Sakstein}. The vertical lines are the bounds proposed in
\cite{Jain}.

First, we consider $n=3/2$ in eq.(\ref{poly}), i.e $\Gamma = 5/3$. This is
the typical polytropic index for low mass stars such as brown dwarfs, with non-relativisitic degenerate electrons. 
In this case, we have plotted in fig.(\ref{mratio3by2}), the variation of $M/M_{Ch}$ as a function of $\Upsilon$ for
various values of the anisotropy parameter. The red, black and blue lines here correspond to $\tau = -0.01$, $0$ and
$0.01$, respectively (the $\tau = 0$ case has been considered in \cite{Saito}). In this case, as $\Upsilon$ is increased
from $-2/3$, we see that the mass ratio decreases up to a minimum value that is obtained for a positive value of $\Upsilon$, 
and then starts to rise with further increase in the ST theory paramter. 

Next, in fig.(\ref{mratio3}), we show a similar
plot for $n=3$, i.e $\Gamma = 4/3$. This corresponds to the equation of state considered by Chandrasekhar, for relativistic 
degenerate electrons without a magnetic field, in white dwarf stars. 
As emphasized before,what we are assuming here is that this equation of state is 
still valid even in the presence of magnetic fields that are below the critical value so that one does not have to invoke Landau 
quantization. 

If we assume that the magnetic field at the core is high, i.e $B \gtrsim 10^{13}~{\rm G}$, then it is possible 
to show that the polytropic equation of state changes to $n=1$, i.e $\Gamma = 2$ \cite{Canuto2}
and that one obtains much higher estimates for the Chandrasekhar mass limit \cite{Bani1}. 
In fig.(\ref{mratio3}), on the other hand, we have used the relativistic equation of state without invoking Dirac quantization, 
and find that without the assumption of high magnetic fields in the core, the ST theory paramter in conjunction with 
a small assumed anisotropy can also push up the Chandrasekhar limit. For example, with $\Upsilon = 1.6$ (which was
the upper bound on $\Upsilon$ proposed in \cite{Sakstein}) and
$\tau = 0.01$, we obtain the mass of the white dwarf as $\sim 2.6 M_{\odot}$, where we have used $M_{Ch} = 1.44 M_{\odot}$. 
Note that for $\tau = 0$, $\Upsilon=1.6$ alone pushes up the stellar mass to about $2.1 M_{\odot}$. 
\begin{figure}[h!]
\centering
\centerline{\includegraphics[scale=0.3]{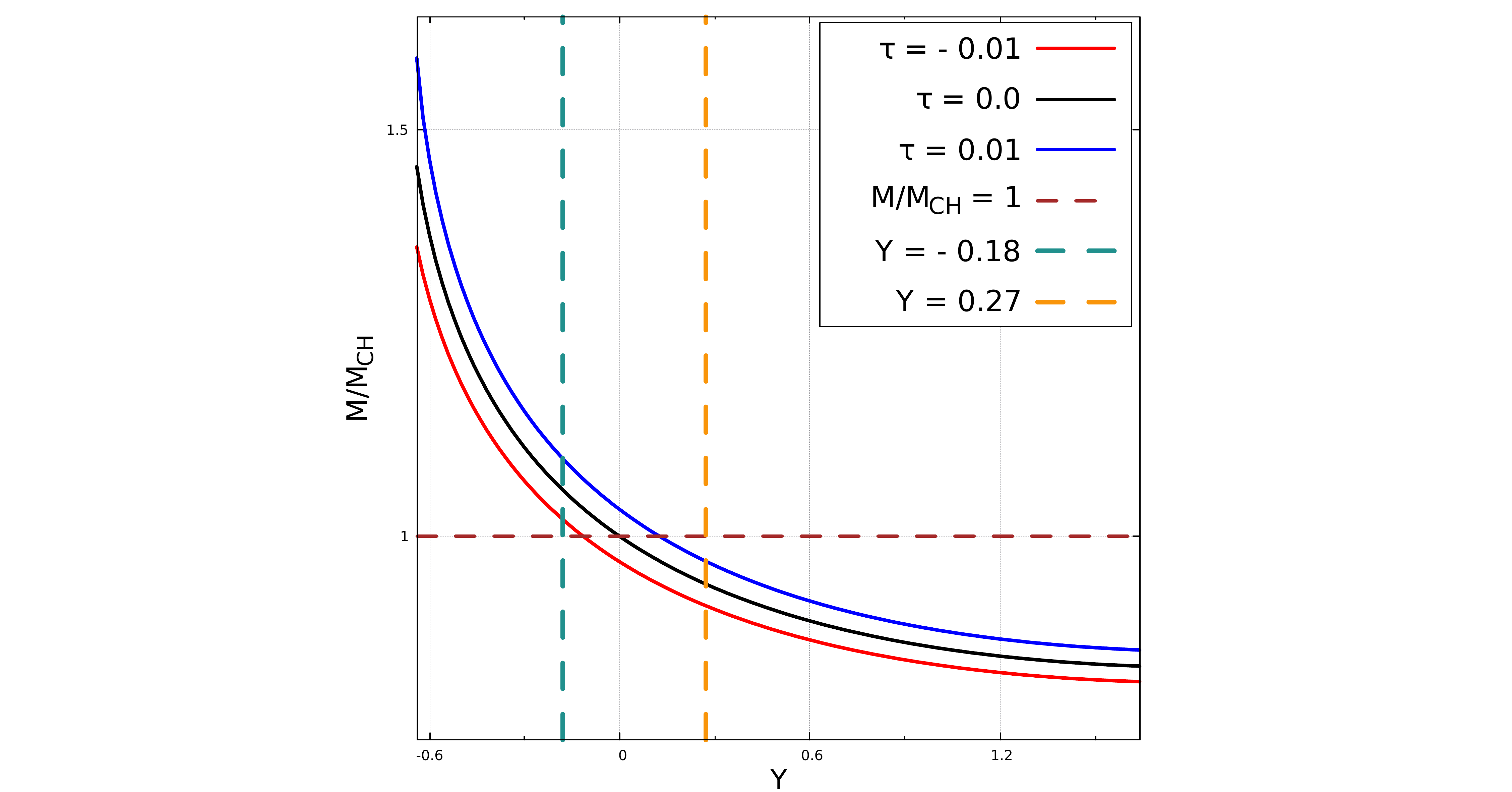}}
\caption{$M/M_{Ch}$ vs $\Upsilon$ for $\Gamma = 2$}
\label{mratio1}
\end{figure}

Next, in fig.(\ref{mratio1}), we plot the same ratio with $\Gamma = 2$, while making the assumption that the
magnetic field does not affect the density and pressures appreciably. To make the assumption more precise, we
recall that \cite{Valyavin} (see also \cite{Parker1} and references therein) in the presence of a magnetic field ${\vec B}$, the
pressure balance equation changes due to the Lorentz force given by $({\vec\nabla}\times{\vec B})\times {\vec B}/(4\pi)$, 
and reads
\begin{equation}
{\vec\nabla}P = \rho{\vec g} -{\vec \nabla}\left(\frac{B^2}{8\pi}\right) + \frac{1}{4\pi}\left({\vec B}.{\vec\nabla}\right).{\vec B} + \cdots ~,
\end{equation}
with ${\vec g}$ being the gravitational force per unit mass, and the ``$\cdots$'' refer to corrections from ST theories and anisotropy. 
In situations where ${\vec\nabla}\times{\vec B}=0$ as  happens for an axial magnetic dipole, the magnetic field terms do not
contribute to the force equation (although pressure anisotropy still remains)
and one can use spherical coordinates to write the hydrostatic equilibrium condition in
eq.(\ref{TOVA}). In general of course there will be deviations from this behaviour, and one will need to add higher magnetic
moments, as discussed earlier. What we are assuming here is that in the latter case also, one can approximately use
eq.(\ref{TOVA}). 

Our numerical analysis indicates that in
this case, increasing the parameter $\Upsilon$ and $\tau$ pulls down the Chandrasekhar mass limit. That this should
be the case is clear from the second relation in eq.(\ref{MCh}), from where it can be seen that increase in $\Upsilon$ decreases
the mass $M$ and hence the mass ratio $M/M_{Ch}$. 
Note that in this case the equation of state implies that the magnetic field at the core is at a value greater than the 
critical value of $4 \times 10^{13}~{\rm G}$, so that quantum effects become prominent, and that only the first Landau
level is filled. Also, in this case, the Chandrasekhar mass limit is obtained by setting $\Upsilon = \tau = 0$ and 
equals $M_{Ch} = 2.58 M_{\odot}$ \cite{Bani1},\cite{Canuto2}.

\section{M-dwarf Scenarios}

Finally, we will make a few brief comments on anisotropy in M-dwarf scenarios. 
Here, in the absence of anisotropy, \cite{Sakstein} solved the Lane-Emden equation (eq.(\ref{LEE})) under some 
reasonable approximations. This was then used to compute the luminosity of Hydrogen burning at the core in low mass stars. On the
other hand, the luminosity of the photosphere was calculated, which also included modified gravity effects, via the second term
in eq.(\ref{TOV}). These two were then equated to relate the minimum mass for Hydrogen burning (MMHB) to the
modified gravity parameter, $\Upsilon$, with $\Upsilon = 0$ giving the results of standard gravity. In this paper, $\theta(\xi)$
was computed near to the center $\xi = 0$ and the resulting solution used for the analysis, which made it analytically
tractable. We will not do such an approximation
here, and resort to a full numerical analysis. 

We will use here \cite{BL} the equation of state for the pressure $P$,
\begin{equation}
P=K\rho^{5/3}~,~K=\frac{(3\pi^2)^{2/3}\hbar^2}{5m_{e}m_{p}^{5/3}\mu_{e}^{5/3}}\left(1+\frac{\alpha}{\eta}\right)~,
\end{equation}
where the degeneracy parameter $\eta$ is defined by
\begin{equation}
 \eta=\frac{(3\pi^2)^{2/3}\hbar^2}{2m_{e}m_{p}^{2/3}k_{B}\mu_{e}^{2/3}}\frac{\rho^{2/3}}{T}~,
\end{equation}
with $T$ being the temperature, $\rho$ the density, and $m_e$ and $m_p$ are the masses of the electron and proton, respectively. 
Also, we use $\alpha = 5\mu_{e}/(2\mu)$, 
where $\mu$ is the mean molecular weight of ionized Hydrogen/Helium mixture, 
and $\mu_{e}$ is the number of baryons per electron.
For H relative abundance $= 0.75$, and He relative abundance $=0.25$, $\mu_{e}=1.143$ and  $\mu=0.593$.

Following the standard treatment given in \cite{BL}, it can be checked that the expression for the 
luminosity due to Hydrogen burning is given, after some straightforward algebra, by
\begin{equation}
L_{HB}= \frac{5.74505\times 10^{7} {M_{-1}}^{11.9733}\omega_{1}\delta^{5.48667}\eta^{10.15}}~,
{\omega_{0} \gamma^{16.46}(\alpha+\eta)^{16.46}}
\end{equation}
where the following definitions have been used : 
\begin{equation}
\omega_{0}=\int_{0}^{\xi_{R}} \xi^2 \theta(\xi)^n d\xi~,~\omega_{1}=\int_{0}^{\xi_{R}} \xi^2 \theta(\xi)^{9.73} d\xi~,~
\delta=\frac{\xi^3}{3\omega_{0}}~,~\gamma=\left[\frac{5\sqrt{10 \omega_{0}}}{16\pi}\right]^{2/3}\xi_{R}~,~
M_{-1}=\frac{M}{0.1M_{\odot}}~.
\label{MDwarf}
\end{equation}
Similarly, the luminosity at the photosphere is given by 
\begin{equation}
L_{e}= \frac{0.5256L_{\odot}\Big(1+\frac{\Upsilon}{2}\Big)^{1.1831}{M_{-1}}^{1.30516}}{\gamma^{0.366197}\eta^{3.98592}
(\alpha +\eta)^{0.366197}}~.
\end{equation}
\begin{figure}[h!]
\centering
\centerline{\includegraphics[scale=0.3]{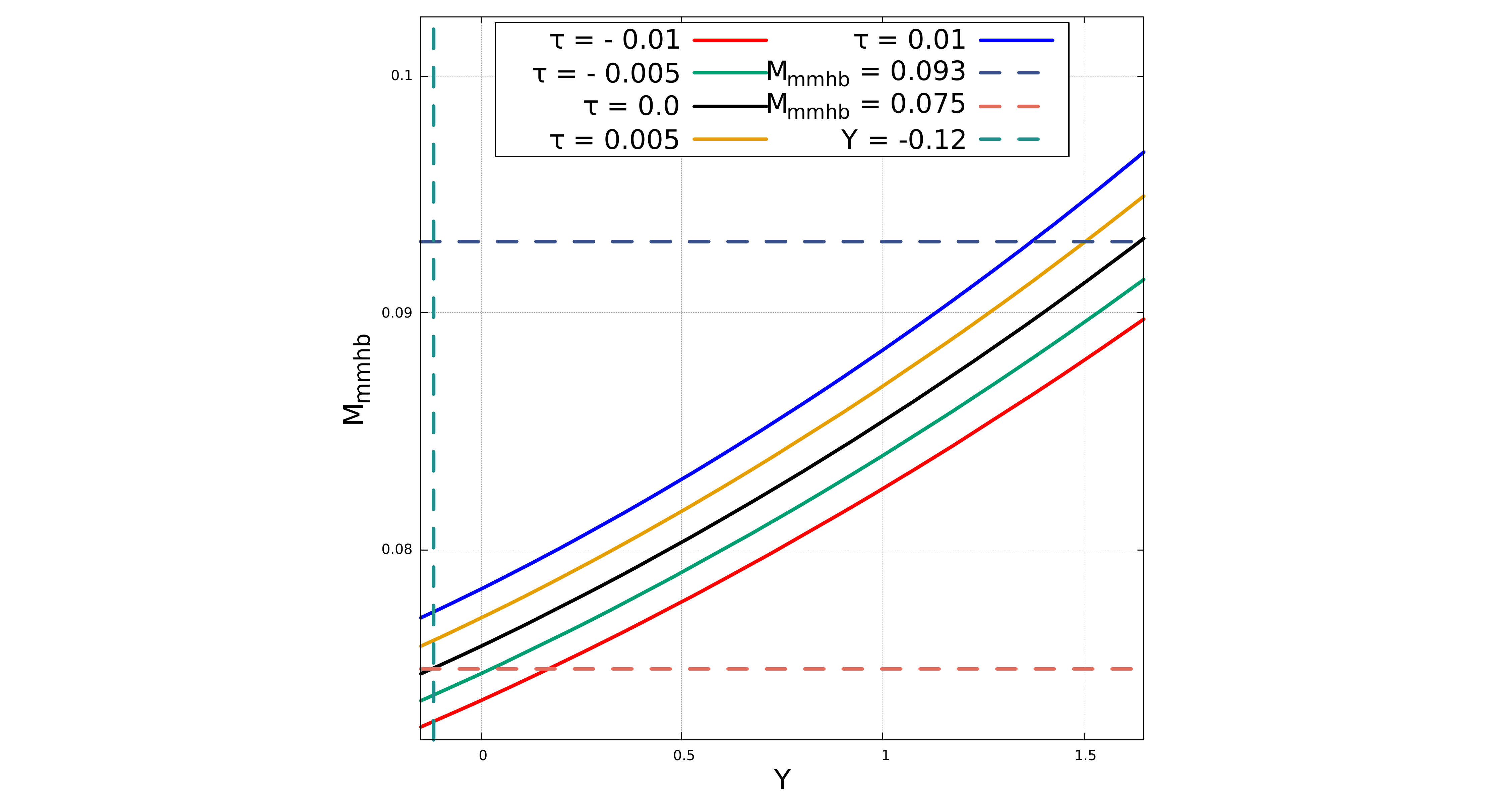}}
\caption{Minimum mass of Hydrogen burning in low mass stars as a function of $\Upsilon$ for different $\tau$, with
$\Gamma = 5/3$.}
\label{Mmhb}
\end{figure}
The final expression for $M_{mmhb}$ that we find is 
\begin{equation}
M_{mmhb}=0.1 M_{\odot}\frac{0.176393\left(1+\frac{\Upsilon}{2}\right)^{0.1109}\omega_{0}^{0.0937365}
\gamma^{1.50858}(\alpha+\eta)^{1.50858}}{\omega_{1}^{0.0937365}\delta^{0.514301}\eta^{1.32505}}~.
\label{finalexp}
\end{equation}
This quantity is plotted for the minimum value of the combination of $\alpha$ and $\eta$ that appears in
eq.(\ref{finalexp}) (given by the numerical value $2.34$) as a function of $\Upsilon$ for different $\tau$ values in 
fig.(\ref{Mmhb}), with the $\tau$ dependence arising out of the solution to the Lane-Emden equation, 
via $\omega_0$ and $\omega_1$ in eq.(\ref{MDwarf}). 

Note that our expression for the MMHB does not match with the one given in \cite{Sakstein} (there appears to be some 
typographical errors in that paper), although
for $\tau = 0$, we do obtain the upper bound mentioned in that paper ($\Upsilon \leq 1.6$) which is the maximum value of $\Upsilon$ in
fig.(\ref{Mmhb}). This is obtained from the known fact that the lowest mass M-dwarf observed has 
$M \sim 0.093 M_{\odot}$ \cite{Segransan}, which is also indicated in the figure. Although fig.(\ref{Mmhb}) might
seem to indicate that with negative values of $\tau$, ST theories that are ruled out in the isotropic case might continue
to be valid in the presence of anisotropy, the fact that such aniotropy can be very small in M-dwarf stars might make this
statement somewhat less significant and only of theoretical interest. 

However, we can make the following observation. First
note that for $\Upsilon = \tau = 0$, our expression is also close to the result $M_{mmhb} \simeq 0.075 M_{\odot}$ \cite{Kumar}, 
\cite{Chabrier} as obtained from a GR computation.
Now, it is generally believed that $M_{mmhb}$ should lie between $0.075 - 0.08 M_{\odot}$. Hence, from fig.(\ref{Mmhb}), 
we get for $\tau \sim 0$, a lower bound $\Upsilon \gtrsim -0.12$ (indicated by the dashed vertical line) 
in order to rule out values of the MMHB below $0.075 M_{\odot}$ in the isotropic case. 
This is close to (but stronger than) the bound $\Upsilon \gtrsim -0.18$ obtained for the isotropic case by Jain et. al. 
in \cite{Jain} from white dwarf scenarios.

\section{Conclusions}

In this paper, we have carried out a comprehensive analysis of equilibrium in stellar matter in 
scalar-tensor gravity, in the presence of anisotropy. We have argued that such anisotropy can modify the stellar
physics in a non-trivial way. In the presence of anisotropy we first presented an algebraic analysis appropriate for 
the physics near the stellar core. Later, we solved the Lane-Emden equations numerically
to arrive at results concerning the mass of the star, and finally revisited the issue of the minimum mass for hydrogen
burning in low mass stars. The main findings of this paper are now summarized.  

\noindent
{\bf 1)} We have shown that in the presence of even a small amount of anisotropy, the bound \cite{Saito} on the ST theory
parameter $\Upsilon$ is modified. That the bound on $\Upsilon$ is dependent on the 
equation of state is also shown. Importantly, this modification leads to an upper bound for $\Upsilon$ (as a function of
the anisotropy) in theories where the
polytropic index is $\Gamma = 4/3$, as is typical for relativistic degenerate electrons in white dwarf stars. \\
\noindent
{\bf 2)} We have seen that for white dwarf stars, ST theories can predict substantially higher values of 
the Chandrasekhar mass, even without the assumption of extreme magnetic fields in the interior, and that this can increase even
further and in a non-linear fashion with the introduction of anisotropy. \\
\noindent
{\bf 3)} In the presence of strong magnetic fields, under the assumption that only the first Landau level is occupied, 
and consequently a modified equation of state with $\Gamma = 2$ should be used, ST theories predict a decrease of the 
Chandrasekhar mass as $\Upsilon$ is increased. \\
\noindent
{\bf 4)} In M-dwarf stars, we show how anisotropy can alter the prediction of the minimum mass of Hydrogen burning.
We show that in the isotropic case, the constraint on the MMHB in M-dwarfs can put a lower bound on $\Upsilon$. This is 
close to, but stronger than the one proposed in \cite{Jain} who arrived at this lower bound via a white dwarf scenario. 

As we have pointed out, the main caveat of our analysis is the assumption of approximate spherical symmetry, which 
prevents us from considering large anisotropies. We have restricted to a particular model of anisotropy due to 
\cite{Heintzmann}, where we have considered the difference between the 
tangential and radial pressures to be much smaller than the core pressure. A (numerical) analysis where such restrictions are not imposed would
be substantially more complicated from the simple minded one presented here, and we hope to report on this in the near
future. 

\begin{center}
\bf{Acknowledgements}
\end{center}

SC thanks Shubhadeep Sadhukhan and Sourav Biswas for code related discussions. TS would like to thank Supratik Banerjee for useful 
comments. We sincerely acknowledge the High Performance Computing
facility (HPC) at IIT Kanpur, India, at which all the numerical computations related to this paper were performed.


\begin{thebibliography}{}
\bibitem{SaksteinPRD} K. Koyama, and J. Sakstein,  Phys. Rev. {\bf D 91}, 124066 (2015).
\bibitem{Saito} R. Saito, D. Yamauchi, S. Mizuno, J. Gleyzes and D. Langlois, JCAP {\bf 1506}, 008 (2015).
\bibitem{Sakstein} J. Sakstein, Phys.\ Rev.\ Lett.\  {\bf 115}, 201101 (2015).
\bibitem{Jain} R. K. Jain, C. Kouvaris, and N. G. Nielsen, Phys. Rev. Lett. {\bf 116}, 151103 (2016).
\bibitem{Holberg} J. B. Holberg, T. D. Oswalt, and M. A. Barstow, Astron. J. {\bf 143}, 68 (2012).
\bibitem{Scalzo} R. A. Scalzo et. al, Ap. J {\bf 713}, 1073 (2010).
\bibitem{MakHarko} M. K. Mak and T. Harko, Proc. Roy. Soc. Lond. {\bf A459}, 393 (2003).
\bibitem{BowersLiang} R. L. Bowers and E. P. T. Liang, Ap. J {\bf 188}, 657 (1974).
\bibitem{HerreraSantosReview} L. Herrera and N. O. Santos, Phys. Rept. {\bf 286}, 53 (1997).
\bibitem{Letelier} P. S. Letelier, Phys. Rev. {\bf D22}, 807 (1980).
\bibitem{Bayin} S. Bayin, Ap. J {\bf 303}, 101 (1985).
\bibitem{ChandraRot} S. Chandrasekhar, MNRAS {\bf 93}, 390 (1933).
\bibitem{KippenWei} R. Kippenhann and A. Weigert, {\tt Stellar Structure and Evolution},
Speinger-Verlag, Berlin (1990). 
\bibitem{Hachisu} I. Hachisu, M. Kato, H. Saito, and K. Nomoto, Ap. J {\bf 744}, 76 (2012).
\bibitem{Bani1} U. Das and B. Mukhopadhyay, Phys. Rev. Lett. {\bf 110}, 071102 (2013). 
\bibitem{Konar} R. Nityananda and S. Konar, Phys. Rev. {\bf D89}, 103017, erratum : Phys. Rev.
{\bf D91}, 029904 (2014).
\bibitem{Browning} M. K. Browning, M. A. Weber, G. Chabrier and A. P. Massey,  Ap. J {\bf 818}, 189 (2016).
\bibitem{BDMag} S. V. Berdyugina, D. M. Harrington et. al,  Ap. J {\bf 847}, 61 (2017).
\bibitem{Canuto1} V. Canuto, and H-Y Chiu, Phys. Rev. {\bf 173}, 1210 (1968). 
\bibitem{Canuto2} V. Canuto and H-Y Chiu, Phys. Rev. {\bf 173}, 1220 (1968). 
\bibitem{Canuto3} V. Canuto, V and H-Y Chiu, Phys. Rev. {\bf 173}, 1229 (1968). 
\bibitem{Ferrer} E. J. Ferrer et. al. Phys. Rev. {\bf C82}, 065802 (2010).
\bibitem{Heintzmann} H. Heintzmann and W. Hillebrandt, Astron. \& Astrophys, {\bf 38}, 51 (1974).
\bibitem{HerreraSantos} L. Herrera and N. O. Santos, Ap. J {\bf 438}, 308 (1995).
\bibitem{HerreraBarreto} L. Herrera, L and W. Barreto, Phys. Rev. {\bf D87}, 087303 (2013).
\bibitem{Madsen} J. Madsen, Ap. J {\bf 367}, 507 (1991).
\bibitem{Shapiro} L. Shapiro and S. A. Teukolsky, {\tt Black Holes, White
Dwarfs, and Neutron Stars: The Physics of Compact Objects}, Wiley, New York (1983).
\bibitem{Chandrasekhar} S. Chandrasekhar, MNRAS {\bf 95}, 207 (1935).
\bibitem{Valyavin} G. Valyavin, O. Kochukhov and N. Piskunov, Astron. \& Astrophys, {\bf 420}, 993 (2004).
\bibitem{Parker1} E. N. Parker, {\tt Conversations on Electric and Magnetic Fleid in the Cosmos}, Princeton
University Press (2007).
\bibitem{BL} A. Burrows and J. Liebert, Rev. Mod. Phys. {\bf 65} (no. 2), 301 (1993). 
\bibitem{Segransan} D. Segransan et. al, Astron. Astrophys. {\bf 364}, 665 (2000).
\bibitem{Kumar} S. S. Kumar, Ap. J {\bf 137}, 1121 (1963).
\bibitem{Chabrier} G. Chabrier and I. Baraffe, Ann. Rev. Astron. Astrophys. {\bf 38}  337-377 (2000).
\end{thebibliography}
\end{document}